\documentclass[a4paper,twocolumn]{esapub}
\usepackage{ifluatex,ifxetex} 
\ifluatex
\usepackage{fontspec}
\else\ifxetex
\usepackage{fontspec}
\else
\usepackage{times}
\fi

\usepackage[numbers]{natbib}
\usepackage{graphicx}

\title{NEARBY Platform for Automatic Asteroids Detection and EURONEAR Surveys}
\author{Dorian Gorgan}
\affil{Technical University of Cluj-Napoca, Cluj-Napoca, Romania, dorian.gorgan@cs.utcluj.ro}
\author{Ovidiu Vaduvescu}
\affil{Isaac Newton Group, Santa Cruz de la Palma, Canary Islands, Spain; 
Instituto de Astrofisica de Canarias, La Laguna, Tenerife, Spain; 
University of Craiova, Craiova, Romania, ovidiu.vaduvescu@gmail.com}
\author{Teodor Stefanut}
\author{Victor Bacu}
\author{Adrian Sabou}
\author{Denisa Copandean Balazs}
\author{Constantin Nandra}
\affil{Technical University of Cluj-Napoca, Cluj-Napoca, Romania, \{teodor.stefanut; victor.bacu; adrian.sabou; denisa.copandean; constantin.nandra\}@cs.utcluj.ro}
\author{Costin Boldea}
\author{Afrodita Boldea}
\author{Marian Predatu}
\author{Viktoria Pinter}
\author{Adrian Stanica}
\affil{University of Craiova, Craiova, Romania, cboldea@inf.ucv.ro, \{alina30.boldea; mariannpredatu; vikii9999; stanicadrn\}@gmail.com}

\begin{document}


\maketitle

\begin{abstract}
The survey of the nearby space and continuous monitoring of the Near Earth Objects (NEOs) and especially Near Earth Asteroids (NEAs) are essential for the future of our planet and should represent a priority for our solar system research and nearby space exploration. More computing power and sophisticated digital tracking algorithms are needed to cope with the larger astronomy imaging cameras dedicated for survey telescopes. The paper presents the NEARBY platform that aims to experiment new algorithms for automatic image reduction, detection and validation of moving objects in astronomical surveys, specifically NEAs. The NEARBY platform has been developed and experimented through a collaborative research work between the Technical University of Cluj-Napoca (UTCN) and the University of Craiova, Romania, using observing infrastructure of the Instituto de Astrofisica de Canarias (IAC), and Isaac Newton Group (ING), La Palma, Spain. The NEARBY platform has been developed and deployed on the UTCN's cloud infrastructure and the acquired images are processed remotely by the astronomers who transfer it from ING through the web interface of the NEARBY platform. The paper analyzes and highlights the main aspects of the NEARBY platform development, and the results and conclusions on the EURONEAR surveys.
\end{abstract}

\section{Introduction}
The survey of the nearby space and continuous monitoring of the Near Earth Objects (NEOs) and especially Near Earth Asteroids (NEAs) are essential for the future of our planet. Therefore, their discovery and monitoring should represent priorities in our solar system research and nearby space exploration. In the past two decades an increasing interest in discovering NEOs has been noted in the astronomical community, but the European contribution remained scarce. More computing power and sophisticated algorithms are needed to cope with the larger astronomy imaging cameras equipped on survey telescopes.

Medium and larger size telescopes (2-4m) are needed for the detection of fainter NEAs using the classic "blink" algorithm aimed to uncover faint targets visible in most individual images. Smaller telescopes could be also used to image faint targets invisible on individual images using the "track and stack" (for known objects) and the new synthetic or digital tracking algorithms (for unknown objects), which need extensive computing resources. This paper proposes improvements to these methods and an approach of integrating them in a new pipeline. This solution assists astronomers to reduce astronomical images and detect moving sources in astronomical surveys of the nearby space in almost real time. The NEARBY Project (Visual Analysis of Multidimensional Astrophysics Data for Moving Objects Detection) \citep{nearby17}, funded by the Romanian Space Agency (ROSA), aims to experiment new algorithms and to develop a software platform to enable automatic image reduction, detection and validation of moving objects in astronomical surveys, specifically NEAs.

The NEARBY platform has been developed and experimented through a collaborative research work between the Technical University of Cluj-Napoca (UTCN) and the University of Craiova, Romania, using observing infrastructure of the Instituto de Astrofisica de Canarias (IAC), and Isaac Newton Group (ING), Spain. The experimental validation has been achieved by survey data provided by the Isaac Newton Telescope (INT) at Roque de los Muchachos astronomical observatory in La Palma. The NEARBY platform is implemented and executed on the cloud infrastructure at UTCN. The acquired images are processed remotely by the astronomers who transfer them from ING through the web interface of the NEARBY platform. The NEARBY platform also supports the visual analysis and validation of the moving objects, and flexible description of the adaptive processing over high performance computation infrastructure. The presentation analyzes and highlights the main aspects of the NEARBY platform development, and the results and conclusions on a few EURONEAR mini-surveys.

This paper is structured as follows: Section 2 presents some related works on the domain of NEA and NEO detection and tracking. Section 3 presents the NEARBY platform used to reduce astronomical images and to detect asteroids. Section 4 describes the asteroids detection algorithm implemented and experimented in the NEARBY platform. Section 5 describes the web graphical user interface that makes available the platform’s functionalities. Section 6 and 7 presents the validation of the platform through the EURONEAR surveys. Section 8 analysis the astrometric accuracy and photometric limits of the NEARBY detections. The last section concludes on the NEARBY platform implementation and validation and sketches some future works.

\section{Related Works}
Since 1980, six US-led NEO sky surveys (Spacewatch, LONEOS, NEAT, LINEAR, CSS and Pan-STARRS) have been using upgraded old or modern 1-2m telescopes which discovered more than 99\% from the entire known NEO population (more than 19,500 objects by the end of 2018). Since 2015 the ATLAS project of the University of Hawaii started a survey using two 0.5m telescopes, which by the end of 2018 has discovered almost 250 NEAs, proving that small telescopes still have a future. In about 5 years, the US-led LSST survey should start to operate  for at least 10 years in Chile, using a 6.5m diam equivalent mirror with the aim to cover the entire visible Southern sky every few nights using short exposures to detect and secure NEOs, besides other science aims.

In the meantime, some pioneering projects in Europe aimed to establish two surveys, namely the ODAS survey at OCA-DLR in France and Germany (1996-1999 which discovered 4 NEAs using a 0.9m telescope) and CINEOS in Italy (2001-2004 which discovered 7 NEAs using a 0.6m telescope). Actually the European NEO survey contributions have been led by two amateur and public outreach facilities, namely the PIKA survey at Crni Vrh in Slovenia (which discovered 27 NEAs since 2003 using one 0.6m telescope) and the LSSS survey in La Sagra, Spain (which discovered about 100 NEAs between 2008 and 2014 using three 0.45m telescopes). Since 2010, ESA uses their 1m ESA-OGS telescope in Tenerife during a few dark nights every month for the NEA survey (TOTAS), discovering 23 NEAs, most of the time being used for tracking and finding space debris \citep{koschny13}, \citep{koschny15a}, and \citep{koschny15b}, which include other similar activities in Europe. Soon, ESA should deploy the "fly-eye" 1m prototype telescope, aiming to increase their NEO and space debris contributions. 

Since 2006, the European Near Earth Asteroids Research (EURONEAR) project aims to increase the European contribution in the NEO field using existing telescopes available in both hemispheres to the members of this network. In particular, it ameliorated orbits of about 1,500 known NEAs based on observations using mostly 1-2m class telescopes (\citep{vaduvescu08}, \citep{vaduvescu11a}, \citep{vaduvescu13a}, \citep{birlan10}) and another 500 NEAs based on data mining of existing image archives of 2-8m class telescopes (\citep{vaduvescu09}, \citep{vaduvescu11b}, \citep{vaduvescu13b}, \citep{vaduvescu17}). This  project involved many students and amateur astronomers who reduced the images and used Astrometrica software to visually search, measure and report all moving objects appearing in all frames, which included few dozen thousands main belt asteroids (MBAs). 9 NEAs were discovered using mostly the 2.5m Isaac Newton Telescope (INT) during about 50 nights total (\citep{vaduvescu15}) and the 2.2m ESO/MPG telescope during only 3 nights, and five NEAs were lost due to lack of telescope time for recovery or late reduction. During two runs (ESO/MPG 2008 and INT 2012) it carried out small surveys in the ecliptic during successive nights (only about 10 sq. deg in total) to visually search for new MBAs and NEAs, discovering a few hundred MBAs. 

Astronomical images are captured using large mosaic cameras, which typically contain an array of CCDs. Initial research on automated asteroid detection systems can be mentioned since 1992, when Scotti et al. \citep{scotti92} conducted a survey for NEAs with a TK2048 CCD in the scanning mode. The image data were transfered to a computer using 3 Sparc CPUs to look for streaked images of nearby asteroids and to identify sets of images that displayed consistent motion. The first synthetic algorithm applied to asteroid detection was developed by Yanagisawa et al. \citep{yanagisawa05}. It combines many images of the same field in order to detect very faint unknown moving objects that are invisible on a single CCD image. 

Well established surveys such as Pan-STARRS \citep{denneau13} or CSS have developed their own moving object processing systems (MOPS) to reduce images, detect moving sources, flag NEO candidates, identify known objects, pair one-night detections in tracklets and multi-night tracklets in unknown objects, calculate new orbits and predict follow-up alerts for newly discovered NEOs.  Other in-house software aimed to smaller surveys are relatively easier to develop today, thanks to the freeware standard astronomical image environments available under Linux such as IRAF \citep{tody86}, SExtractor and Astromatic \citep{bertin96}, \citep{bertin06} and Astrometry.net \citep{lang10}. Other stand-alone codes for field recognition, moving object detection and astrometric reduction have been written by dedicated amateur astronomers and computer developers under Windows (mostly in C), such as Astrometrica \citep{astrometrica18}, SkySift \citep{holvorcem15}, PinPoint \citep{pinpoint18} and by small teams under Linux \citep{allekotte13}. 

Copandean et al. \citep{copandean17}, \citep{copandean18} propose an automated pipeline prototype for asteroids detection written in Python under Linux, which calls some 3rd party astrophysics libraries. The current version of the proposed pipeline prototype is tightly coupled with data obtained from the 2.5 meters diameter Isaac Newton Telescope (INT) located in La Palma, Canary Islands, Spain and represents the basis on which the NEARBY platform is built upon.

\section{Nearby Platform}
The design of the NEARBY platform started by having in mind the following basic objectives: 
\begin{itemize}
\item Process astronomical images to detect and identify moving objects (asteroids);
\item Possibility to dynamically scale the system in order to accommodate the processing of huge volume of data;
\item Integrate human assisted validation of the detections by visual analysis techniques;
\item Possibility to modify the processing pipeline by adding other functionalities;
\item Deploy the NEARBY platform on a cloud based  computing infrastructure.
\end{itemize}

\begin{figure}
\centering
\includegraphics[width=1.0\linewidth]{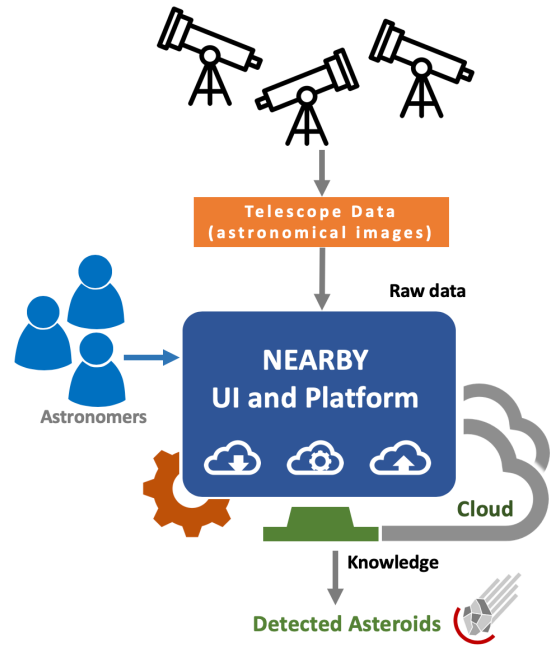}
\caption{Conceptual NEARBY architecture.}
\label{fig:arch}
\end{figure}

Client-server applications, such as web  applications, are often built using a 3-tier architectural pattern. NEARBY platform is composed of three layers or tiers, namely the presentation, logic and data layers, each one being built on top of the others. This kind of architecture comes with some benefits such as modularization, flexibility, scalability, performance and availability. Being conceived as a modularized architecture the product development is faster and very flexible for adding more features or modifying existing implementation of some of the modules.  Figure~\ref{fig:arch} highlights these layers.

The data layer supports storage of all the information needed for both automatic execution and management of the asteroids detection (in order to validate or invalidate them by human reducer). It relies on MySQL server for storing metadata related to surveys and asteroid detections in a relational database. Data is available to the logic layer via API calls and is exposed to users through the graphical user interface. The actual data (astronomical images and catalogs) are stored in a directory structure that maps logically each survey data. A survey groups a collection of astronomical images captured with the same telescope during a continuous time interval (several observing nights). In each observing night several fields are observed and this data (images) representing the raw input data (in FITS format) is stored in this hierarchical directory structure.

The logic layer that drives application's core capabilities is built in Python and exposes all the functionalities as REST web services. The functionalities are grouped in three modules: data, execution and validation. The data module deals with survey data management and persisting data to the database. The execution module enables processing of survey data (i.e. image reduction, field correction, source detection and automatic detection of asteroids). The validation module allows human validation of the detections.

\begin{figure}
\centering
\includegraphics[width=0.7\linewidth]{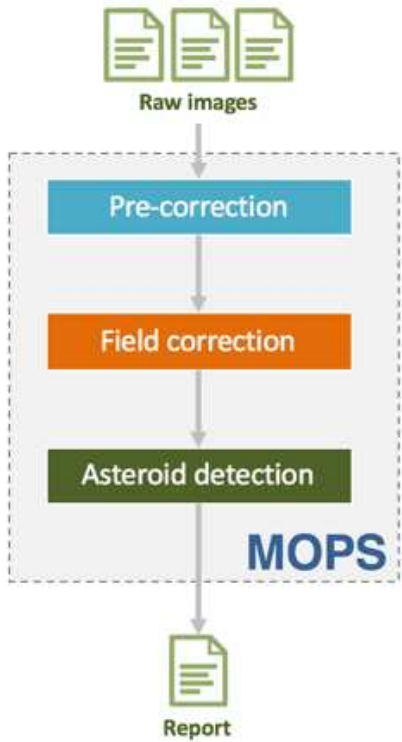}
\caption{NEARBY implements the MOPS concept.}
\label{fig:mops}
\end{figure}

The presentation layer exposes all the NEARBY platform functionality in an intuitive and flexible way. It is developed using HTML5 and Bootstrap and is accessible  through web browsers. This is one advantage of NEARBY compared with other similar application (which are mostly desktop based applications). It supports multiple users to simultaneously work on multiple astronomical images in order to detect asteroids, or to collaborate on reducing and validating potential asteroids easier. Each of the logic layer modules has a presentation layer equivalent. The data management UI supports users to upload surveys (raw images), to modify configuration files and to download results. A dedicated UI for validating asteroids detections allows users to visualize an animated sequence of images of asteroid detection and to validate/invalidate each detection.

The infrastructure supporting this architectural pattern is based on Kubernetes and Docker containers. It encapsulates every layer inside different containers, making it more easily to change and to adapt the configuration to a particular use case. Containers provide resource isolation and allocation by virtualizing the operating system. A Docker container image can be seen as a executable package of software including all the pieces needed to run an application. NEARBY platform consists of various container images for all the layers.

NEARBY platform is powered by a testing environment with the following resources: 3 IBM Blade servers, Intel Xeon Processor E5-2630 v2 6C and 64 GB memory, and 1 IBM Blade server, Intel Xeon Processor E5-2630 v2 6C and 16 GB memory. On top of these computational resources is built a cloud infrastructure running the free and open-source software platform OpenStack. A Kubernetes cluster that manages the containerized applications runs inside this cloud infrastructure.

NEARBY uses MOPS implemented by combining different third-party applications aimed to identify potential asteroids in a sequence of astronomical images targeting the same field. The raw images (a sequence of at least three images) are processed sequentially (Figure~\ref{fig:mops}) by three modules (pre-correction, field correction and asteroid detection). 

The pre-correction module receives as input the set of raw images and corrects artefacts and bad pixels from them by using flat, bias and badpixel images. This module is written in Python and uses IRAF to apply the flat and bias images after combining them in bias and flat master images. The output of this module is a set of processed images.

The field correction module reduces the field distortions and resamples the set of images corrected by the previous module. This module is also written in Python and uses several third-party applications such as IRAF (correct headers), SExtractor (extract a list of sources), SCAMP (compute shifting function used to correct field distortion), and SWARP (resample images based on the shifting functions).

The asteroid detection module (which will be presented in more detail in a later section) identifies asteroids trajectories from the sources identified by SExtractor. All the detections rely just on the captured images and the algorithm implemented in this module is based on the assumption that an asteroid moves with constant speed and on a trajectory that can be considered linear (during the short observation window).

\section{Asteroids Detection Algorithm}
As mentioned in the previous section, the detection algorithm relies only on the astronomical objects identified by SExtractor from the analysed images, which can be stars, noise or asteroids. The SExtractor output catalogues include, for each of the identified objects, information like: approximate position in RA and DEC coordinates, apparent magnitude, ellipticity form etc. Each object is catalogued as a line of text within an ASCII file, which contains only the required elements from a single analysed image. As the analysis performed by SExtractor is independent for each individual image, no correlations or context information is included for the astronomical objects.

As a result, at the start of the algorithm we have no information on the nature of each object (if it is a star, a galaxy, noise or asteroid) and no indication if a specific object from one image changed its position in the next ones or if it is even visible in other images. In theory, at this point any of the identified astronomical objects could be a potential asteroid. In order to identify the actual candidates we will assume that asteroids are moving on a linear trajectory in the observation time interval, with a constant speed.

The algorithm needs the following information in order to extract only valid asteroid candidates from the available SExtractor catalogs (objects pool):
\begin{itemize}
\item observing and exposure time of each image (in Julian Day format);
\item set of parameters that describe specific attributes for the survey and the asteroids of interest;
\item set of ASCII generated SExtractor catalogs with the identified astronomical objects.
\end{itemize}

The input parameters of the algorithm are:
\begin{itemize}
\item \textit{pixel scale} - the number of arc seconds covered by each pixel in the image;
\item \textit{minimum and maximum speed} ($\mu _{min}, \mu _{max}$) - the speed limits measured in arc seconds/minute for the moving objects of interest. This value should be coordinated with the survey location on the sky and with the time interval between two consecutive images of the same field;
\item \textit{maximum allowed speed tolerance} ($\Delta \mu _{max}$) - the maximum variation that is accepted between different parts of the same trajectory (influenced mainly by positioning error);
\item \textit{maximum allowed positional tolerance} ($\epsilon _{max}$) - the maximum distance between two objects identified in different images in order to be considered instances of the same fixed entity (stars and galaxies).
\end{itemize}

Based on the information received as input, the algorithm will infer all the contextual connections required to link all the instances of the same object identified in separate images, using the following main steps (which have been detailed in \citep{stefanut18}):

\textit{1. Identify and remove fixed objects from the objects pool}
If in at least two different images there is an object placed at the same position within positional tolerance, then it is considered a static element (star, galaxy or noise), and will not consider it further as an asteroid candidate. In this phase a major role is played by $\epsilon _{max}$ which will determine if two objects placed very closely to a certain position are treated as instances of the same entity or as independent entities.

\textit{2. Split the valid speed interval in smaller intervals}
Typically NEOs have an apparent movement speed of 0.05 to 10 "/min. This large variation favours the generation of many false positives trajectories, because each of the astronomical candidate objects can be paired in multiple apparently valid trajectories, with different movement speeds. This problem is especially noticeable in the initialization phase where the apparent speed is one of the most important validation factors. In order to address this issue, we are running the following phases of the algorithm multiple times, by splitting the entire domain ($\mu _{min}, \mu _{max}$) in 0.5 "/min intervals.

\textit{3. Initialize possible trajectories} 
For each of the speed intervals defined at the previous step, we will attempt to identify all the potentially valid trajectories. The first instance of each asteroid candidate will be chosen from the clearest image available in the analysed set, based on the astronomical seeing conditions (FWHM) which is one of the parameters retrieved from SExtractor catalogues. This ensures that the number of possible asteroid detections is maximized, as the chances for the asteroids being visible in this image are the highest.

The analysed object is considered as a potential asteroid if and only if, when paired with all the objects from an adjacent image based on the apparent velocity, only one valid pair can be created. This pair will also represent the initial segment of a future trajectory, that will be further developed in the next phase.

\textit{4. Trajectories development} 
From each of the remaining images of the analysed set we are selecting a single astronomical object for every trajectory initiated in the previous phase. Any selected object can be included in a single trajectory, the choice being made according to the cost function presented in (\ref{eq:eqf}). It is important to underline the fact that applied conditions cannot be very restrictive because positioning errors, caused by the astrometric solution, are very likely to occur. In our tests, the maximum positioning error identified had a value of 0.7".

\begin{equation} \label{eq:eqf}
f = 0.5 * (1 - \frac{\mid \mu _{t} - \mu _{c} \mid }{\Delta \mu _{max}}) + 0.5 * (1 - \frac{\mid \theta _{t} - \theta _{c} \mid }{\Delta \theta _{max}})
\end{equation}

where:
\\$\mu _{t}$ - the average speed on the existing trajectory segments;
\\$\mu _{c}$ - the speed computed on the candidate segment for this trajectory, if the current object would be selected for inclusion in the trajectory;
\\$\Delta \mu _{max}$ - the value of the \textit{maximum allowed speed variation} parameter;
\\$\theta _{t}$ - the average angle between the existing trajectory segments and the chosen reference;
\\$\theta _{c}$ - the angle of the candidate segment for this trajectory, if the current object would be selected for inclusion in the trajectory;
\\$\Delta \theta _{max}$ - the maximum deviation angle accepted when searching for candidates, computed dynamically as presented in (\ref{eq:eqalfa}).

\begin{equation} \label{eq:eqalfa}
\Delta \theta _{max} = \arcsin {\frac{2* \epsilon _{max}}{L _{t}}}
\end{equation}

where:
\\$\epsilon _{max}$ - the value of the \textit{maximum allowed positional tolerance} parameter;
\\$L _{t}$ - the length of the already defined trajectory.

The output of the described algorithm has two main components: (1) identified trajectories in MPC format \citep{mpc18} and (2) image-space coordinates for each of the identified NEA candidates. This data will be further used by the NEARBY UI component to display relevant information to the human validator, as described in the following sections.

\section{Asteroids Detection Application}
 The NEARBY platform provides to astronomers a complete set of tools that enable them to easily analyse large amounts of astronomical images with the purpose of identifying NEAs and other asteroids. The application includes the necessary functionalities for managing telescope specific information (ex. flat and bias images), astronomical actual data (series of astronomical images), the configuration files of different libraries (ex. IRAF, SExtractor, Scamp, etc.) as well as the output information defining the automatically detected asteroids.
 
All the features are exposed to the user through the graphical interface of the NEARBY UI module, which has been developed as a web application. By authenticating with a username and a password, any astronomer interested in searching for NEA candidates using his own astronomical images can access the available functionalities.

\subsection{Data Management Model}
Inspired by the current observing techniques, the application presents to the specialists a familiar data management model that reduces the learning time and streamlines the configuration process. The information is organized in a tree-like fashion, with the following levels:

\subsubsection{Experiment Level}
An experiment represents the top-level of the hierarchy and is equivalent with an entire survey performed with a telescope for any number of consecutive nights known as "observing run". At this level the specialists have the possibility to input any telescope specific data that should be used in processing of the astronomical images captured during the nights included in the experiment. Any user can create any number of experiments.

\begin{figure}
\centering
\includegraphics[width=1.0\linewidth]{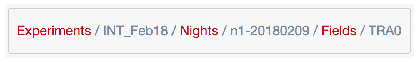}
\caption{Input data visualisation.}
\label{fig:datamodel}
\end{figure}

\subsubsection{Night Level}
This notion describes an actual observation night. Usually the surveys performed with a specific telescope span a few consecutive nights, that should be grouped together under the same \textit{Experiment}. At this level, the user can input all the data and processing parameters that are applicable for an entire night of observation, like the \textit{bias} and \textit{flat} files or the configuration parameters of the different NEARBY libraries. A night entry can contain any number of fields.

\subsubsection{Field Level}
It is a data level designed to map an actual observed astronomical field. Here, the user can upload the sequence of astronomical images that has been captured and needs to be processed for NEA discovery. 

\begin{figure*}
\centering
\includegraphics[width=0.95\linewidth]{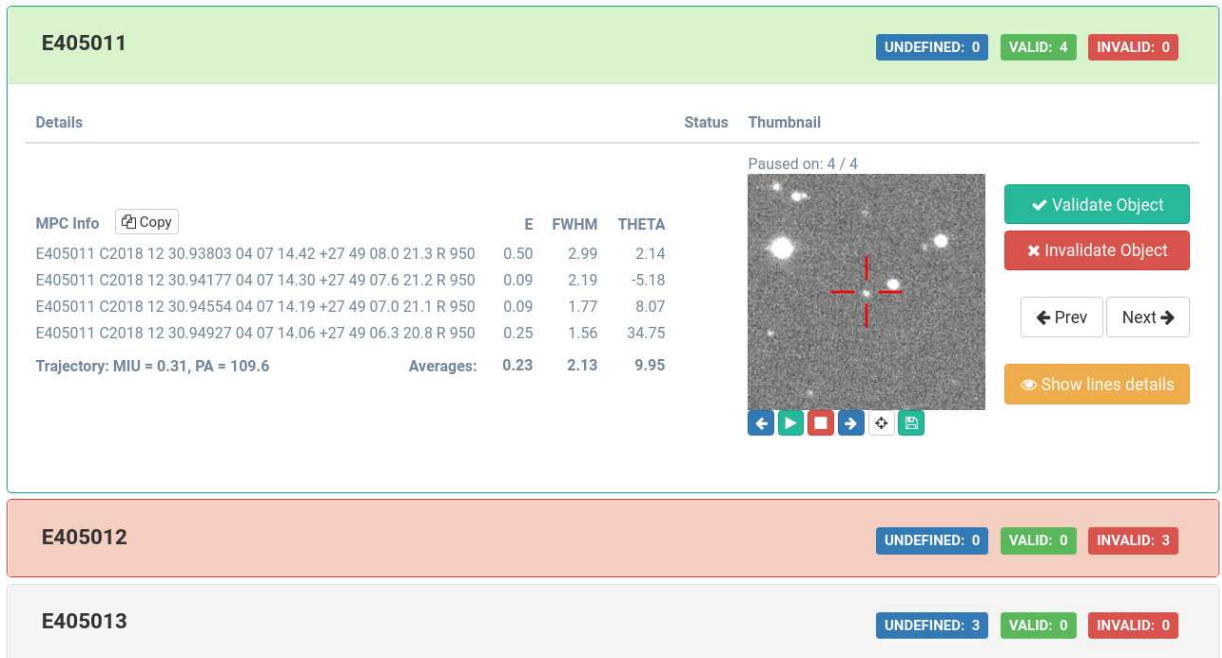}
\caption{Output data visualisation.}
\label{fig:gui}
\end{figure*}

\subsubsection{Project Level}
Represents the last and most detailed level in the data management structure, and has been created to model a specific processing that is performed on the field's data. A project is usually defined by the set of input parameters - telescope specific information, bias and flat files, configuration files of the processing libraries, astronomical images of the field - and the final result of the processing: the list of automatically identified moving objects and NEA candidates. The user can create any number of projects for a certain field.

The purpose of this level is to add flexibility to the processing flow while minimizing the data management overhead. By default, all the input parameters of a project are inherited from the upper levels: experiment, night and field. However, when considered necessary, the user has the possibility to overwrite only those default settings which are not optimal for this specific field - for example when the seeing conditions are changing during the same observing night.

\subsubsection{Input Data Visualisation Level}
At all times, the application's interface will display to the user the information on the current visible data, under the form of a hierarchical path. In this manner, the user can easily establish what input parameters defined in upper levels are influencing the processing of a certain field.

In the example presented in Figure~\ref{fig:datamodel}, the user is currently viewing the Projects list of the \textit{Field} TRA0 from the \textit{Night} n1-20180209 of the \textit{Experiment} INT\_ Feb18.

In order to navigate to the Nights list of the \textit{Experiment} INT\_ Feb18, the user can simply click on the \textit{Nights} link. Similarly, if the user would like to switch directly to another Experiment, all that is required is to click on the \textit{Experiments} link and then select the desired entry from the list.

\subsection{Data Processing}
After all the input data has been uploaded into the NEARBY platform, the user can launch the processing by simply pressing the Run button next to the desired project. The platform will automatically schedule the processing and launch it as soon as the necessary resources are available, without any other explicit intervention from the user.
When the processing has been successfully completed, two new functionalities will become available for the project:
\begin{itemize}
\item \textit{MPC Report} - allows the user to view all the asteroid candidates automatically identified by the NEARBY platform and to validate if they should be included or not in the final MPC Report
\item \textit{download results} - the user can access the important intermediate results of the processing in raw format (catalogs, plots, reduced images)  
\end{itemize}

\subsection{Output Data Validation}
Once the processing of a specific project has been completed, the user can access the complete list of asteroid candidates automatically identified by the NEARBY platform using the algorithm described in the previous section. This list might contain false positives and needs to be manually validated by an assistant reducer in order to generate the final MPC Report with the valid identified asteroids.

Depending on the seeing conditions and on the values of the processing parameters, the number of total results can reach dozens or even hundreds of detections in each field. When searching for NEAs, time is essential. Ideally, best NEA candidates can be followed in the same night or, the latest, in the following one.

It is possible the optimizing of the validation process only by reducing the number of automatically generated results through parameters fine-tuning, but it often requires multiple trials (so more time) and, from a certain point onwards, can lead to the elimination of very good candidates (for example fainter objects). Also it is very important to reduce the time necessary for validation as much as possible, in order to reduce in the real time and to maximize the chances of NEA discovery.

Addressing these principles, the NEARBY application displays initially only the essential information for each object, which is typically enough for the astronomer to take an informed decision. As can be seen in Figure~\ref{fig:gui}, for the object E405011, the user can visualise the main trajectory parameters in a table format and the movement animation created by the platform from the captured images. The animation can be controlled (play, pause, stop) and played frame by frame, enabling the user to thoroughly analyse the optical data in a manner very similar to the blinking method, which is widely used by the astronomical community.

Validated objects are highlighted in green colour while the rejected ones are coloured in red. The objects that have not yet been evaluated are displayed in grey. In this manner, the user can easily verify the status of an object without the need to visualise its data.

For more complex situations, the user can use the \textit \textit{Show line details} functionality (Figure~\ref{fig:gui}) to display and validate each identified position for the object. In situations when some of the astronomical images in the set are completely or partially damaged (bad seeing conditions, telecamera hardware error etc.) it is possible that valid asteroid trajectories to result from noise or positioning errors (in only one or two images). In this cases, the validator has the possibility to remove anytime those positions from the final MPC report while validating the rest of the trajectory. The status of these partial validations can be easily verified by the user through the three badges (blue, green and red) displayed next to the object's name. The numbers displayed there show how many positions have been validated, invalidated or have not yet been evaluated.

In most of the situations, the user has to perform only two explicit interactions with the interface in order to evaluate an object and to mark it accordingly: (1) press the Validate Object or Invalidate Object button (depending on the personal decision); (2) press the Next button to hide the information of the current object and display the information for the next. 

\section{EURONEAR and NEARBY Pilot Surveys}
The 2.5m Isaac Newton Telescope (INT) is hosted by the Isaac Newton Group, being installed at 2300m altitude at Roque de los Muchachos Observatory (ORM) in La Palma, Canary Islands, Spain. At its F/3.3 prime focus it is equipped with the Wide Field Camera (WFC), a mosaic consisting in four 2kx4k pixels CCDs with pixel size 0.33" covering together 0.27 sq. deg on sky arranged in an 34'x34' L-shape design with about 1' gaps between CCDs. 

Since 2009 the EURONEAR project used the INT/WFC facility for recovery of known NEAs in need of orbital improvement, and during 2013-2016 it conducted a series of override programs to target few hundred one-opposition NEAs \citep{vaduvescu18}. This work engaged a remote team of students and amateur astronomers who visually blinked images in Astrometrica to search and measure all moving objects in all fields including up to ~50 asteroids (about half known and half unknown) appearing in each WFC field observed towards opposition. For each field the image processing was done using the GUI version of THELI \citep{erben05}, \citep{schirmer13}, followed by an IRAF script to make the output Astrometrica-ready, and this process took about one hour, usually being accomplished next day. The observing runs were sparse and short, most of them conducted in override mode accessing only one hour during the nights when the faint NEA targets became feasible (mag 22-23), thus the data reduction load was appropriate for human work who took as much as two hours to carefully reduce in Astrometrica each observed field. Part of the observed fields, in 2014-2015 we serendipitously discovered and secured the first 9 NEAs using the INT - first such discoveries from La Palma, while other 3 NEAs were lost due to the lack of telescope time for recovery or bad weather \citep{vaduvescu15}.

Since 2016 we have continued to build on the MOPS reduction pipeline \citep{copandean17}, \citep{copandean18}. In 2017-2018 this work could be integrated in the NEARBY platform to support some NEA and other asteroid surveys and cover larger sky areas in much shorter time using the INT or other larger survey telescopes. The need of human reduction has been decreased, so that the assistant reducers only need to validate the NEARBY automated source detections and perform some quality control check for each validated field, a process which takes about half hour per WFC field, and in the near future we plan to minimize further the human interaction needs. 

During 2018 we tested NEARBY in near real time using the INT/WFC during few nights (hours) in May and September, and in November we had the first observing run (5 nights). Next we detail the setup and findings of these NEARBY tests and NEA pilot mini-surveys. 

\section{Case Studies Execution}
Covering faster and deeper large sky areas is essential for NEA survey work, and the INT aperture combined with the good ORM seeing are great assets. Although WFC camera was considered wide in 1997 when saw first light, today there are much larger cameras due to wider optics designs. Thus, the cadence should be carefully accounted for any survey with the INT/WFC. 

\begin{figure*}
\centering
\includegraphics[width=1.0\linewidth]{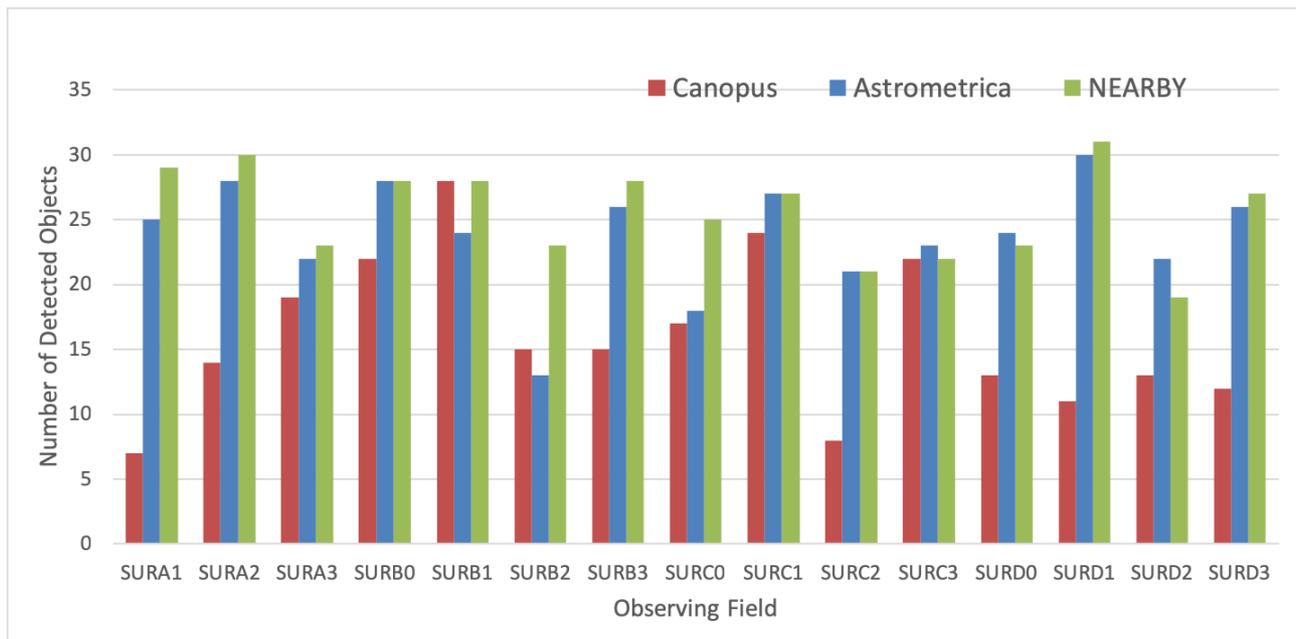}
\caption{Comparing NEARBY with automatic Astrometrica and Canopus over the NEASUR 2017 data pack.}
\label{fig:tool}
\end{figure*}

\subsection{INT/WFC Survey Time Budget}
The next sections analysis the parameters involved in the time budget and NEA discovery rate.

\subsubsection{Exposure Time}
In typical INT seeing 1.2" (historic average), using exposures of 20-30s in dark time and average airmass 1.2, the WFC camera could detect magnitude V$\sim$22.5 at S/N=4-5, which are appropriate for discovery of faint NEAs. The INT tracking produces good images up to 2-3 minutes exposure time, while guiding star acquisition is quite slow and improper for surveys, so we always used tracking in our entire EURONEAR and NEARBY work. 

Shorter exposures are better for minimizing the trail loss effect of faster NEAs and also for increasing the survey sky area. Small trails (up to about 3 times the average seeing, thus 4" long) could be detected and paired by NEARBY (which is not equipped yet with any trail detection algorithm), thus 30s exposures could detect NEAs as fast as 8"/min, while 20s exposures could detect brighter NEAs as fast as 12"/min. 

\subsubsection{Binning and Readout}
Fast WFC readout works with similar results as slow readout (both having similar noise), and the fast readout reads the whole mosaic in 29s (versus 48s the slow mode), so we always used fast readout in our entire survey work. 

The WFC pixel scale is 0.33", relatively small for NEA surveys but providing accurate astrometry and photometry, and during our entire work we used default binning 1. Binning 2 mode is also available (with 25s readout in slow and 15s in fast) and doubles the pixel to 0.66" (better for NEA surveys), but the noise is higher (by about 1.5 times), and much higher in CCD2 (showing some interference pattern which makes CCD2 actually inappropriate). Binning 2 and fast readout could be eventually used to decrease the readout to 15s, at the price of dropping surface coverage by one quarter due to the loss of CCD2 (not preferable). 

\subsubsection{Number of Repetitions}
Minimum 3 repetitions per field are needed for moving objects pairing in very short time, but only three are vulnerable to loss of faint objects (which might not be detected in all 3 images) and also miss-pairing due to noise (especially in poor seeing) which result in more time spent to reject artifacts during validation. More than 4 repetitions increase the pairing confidence but consume more time, while more than 5 are not recommended for a slow camera, thus we decided to use mostly 4 repetitions in our pilot surveys. 

Assuming only 4 repetitions per field, exposures of 30s in fast mode, 29s readout time and about 20s telescope slew time from one field to its neighbor, the 4-point sequence (4 neighboring fields ABCD) will result in a cadence of 5.3 minutes between two consecutive images of the same field, 16 minutes orbital arc and will take 21 minutes to execute. The 9-point sequence (9 neighboring fields ABCDEFGHI used for slower moving objects) will result in 12 min between two consecutive images of the same field, 36 minutes orbital arc and will take 48 minutes total to execute. This seems more appropriate for survey work but increases the risk for fast moving NEAs to exit the relatively small WFC field during the entire sequence (e.g. an NEA with $\mu$ =10"/min will move by 6' during the 9-point sequence). 

\subsubsection{INT/WFC Sky Coverage and Discovery Rate}
Using the WFC fast readout in binning 1 and assuming 30s exposure time, the INT/WFC could cover 100 fields (about 25 sq. deg) to depth V=22.5 in dark time during the average 9 hour long night. In comparison, the well established Pan-STARRS1 survey covers 7 sq. deg each field and 1000 sq. deg each night (142 fields), producing up to $\sim$20 NEA discoveries every night. Scaling down these PS1 numbers (because of the similar apertures and sites), the INT/WFC should be able to discover one NEA every two nights, and this rate actually matches exactly the results of our November 2018 observing run. 

\begin{figure*}
\centering
\includegraphics[width=0.95\linewidth]{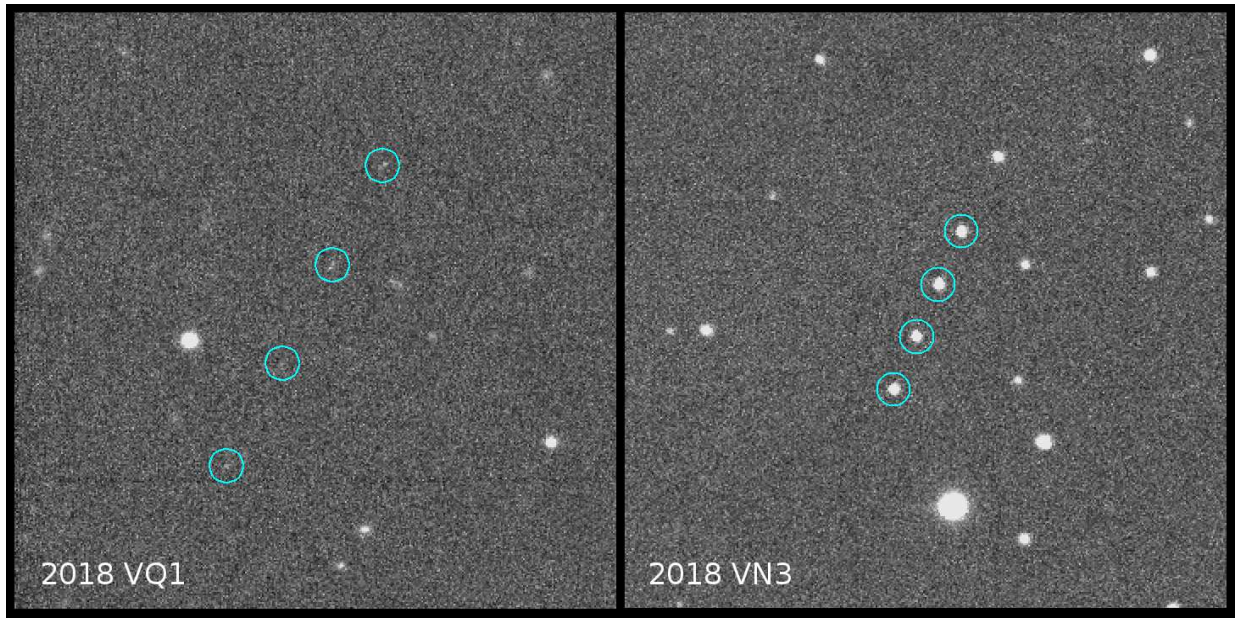}
\caption{First two NEAs (composite images) discovered with the NEARBY platform during the INT observing run in Nov 2018. The sky orientation is normal (N up, E left) and field of view is  2'x2'. The fainter 2018 VQ1 moved upwards, while the brighter 2018 VN3 moved downwards.}
\label{fig:discov}
\end{figure*}

\subsection{Comparison with Other Detection Software}
During the 27 March 2017 night (3 hours morning dark time) it was used the INT/WFC to acquire 16 WFC fields along the ecliptic (4 sq. deg), aiming for the first tests of the MOPS pipeline developed through a doctoral research. 60s exposures were used to observe the 4-ppoint sequence with 4 repetitions. Next day we reduced the run using THELI, IRAF and Astrometrica, involving a team of reducers who carefully blinked images to detect and measure all moving sources. Later this archival dataset was reduced in NEARBY, in the aim to compare results with other software and human detection. 

First, the NEARBY detections of each field were compared with human detections in Astrometrica visual blinking mode. Overall, the NEARBY tops about 90\% results from human detections, with a few cases of asteroids lost by humans but found by NEARBY. 

Second, comparing NEARBY with other two software tools such as Astrometrica in automatic detection mode, and Canopus \citep{warner18} over the NEASUR 2017 data pack, the following results have been obtained: NEARBY detected 384 valid asteroids compared to Canopus that detected 240 asteroids and Astrometrica 357 in automatic mode from a total of 430 manually identified by human reducers (Figure~\ref{fig:tool}). Overall, NEARBY performs better by about 10-20\% than Astrometrica. 

\subsection{NEARBY Tests}
During three nights 2-4 May 2018 (6 hours evening dark time), INT-WFC was used to acquire 56 fields along the ecliptic (about 15 sq. deg total) in the 4-point sequence and 4 repetitions of each field (ABCD-ABCD-ABCD-ABCD). The slow proper motion of most main belt asteroids ($\mu$ $\sim$0.1"/min) in the observed fields combined with the 4-point sequence and bad seeing (2-3") confused NEARBY which detected many artifacts (stars, galaxies) besides relatively few asteroids. 

During two nights 15-16 Sep 2018 (6 hours morning dark time), INT-WFC was used to cover 36 WFC fields (10 sq. deg). It repeated 5 times the 9-point observing sequence (ABCDEFGHI- ABCDEFGHI- ABCDEFGHI- ABCDEFGHI- ABCDEFGHI) which allows more time for the slow objects to appear more distant in neighboring images. The increased number of repetitions did not improve the detection rate, the slower cadence and good seeing remaining essential for the separation of moving asteroids from fixed objects (stars and distant galaxies). 

\subsection{NEARBY Mini-Survey}
During five nights 31 Oct - 5 Nov 2018 (full nights, dark and some gray time) the first NEARBY survey (observing run C10) was conducted, during which 355 survey fields were covered in 38 hours (about 100 sq. deg). Each night a few random fields were planned along the ecliptic, avoiding large surveys (based on MPC Sky Coverage plots). To maximize detections, higher fields were observed in the sky in three parts of the ecliptic, using 4-point sequences for faster moving areas and 9-point sequences for slower areas. To maximize sky coverage and to minimize trailing loss, 30s integration time was used. 

One INT observer was responsible for sorting and uploading the images to the NEARBY server, which usually takes 2-3 minutes for each field (4 WFC raw FITS images adding together about 300 MB) at normal internet speed ($\sim$2 MB/s). Once the observing sequence is finished, the observer could start the simultaneous execution of 4 or 9 fields. 

Each NEARBY run of one WFC field typically takes 10-15 minutes in normal conditions, and a few fields could be run simultaneously, thanks to the NEARBY cloud architecture.

The source validation was distributed based on fields to a remote team of $\sim$15 assistant reducers assigned via a Google Drive live spreadsheet by the run coordinator. Once a field finishes running, it is allocated to one remote assistant reducer who validates the automated detections.

Using the NEARBY web-based interface, each field typically takes 5-10 minutes for the reducer assistant to validate detections (up to about 50 moving sources in WFC ecliptic fields). Following validation, each reducer is responsible for identifying all known asteroids in his/her field using the AsterID tool temporarily deployed on the EURONEAR website (http://www.euronear.org/tools/AsterID.php) which queries the SkyBoT service \citep{berthier06}. Then, each reducer tests the entire report against the MPC NEO Rating tool and the EURONEAR NEA Checker tool (http://www.euronear.org/tools/NEACheck.php) to alert for any possible NEA candidates. The images reduced by NEARBY are then downloaded and visually quick-scanned by the reducer in Astrometrica to spot any possible trail not detected by NEARBY. Finally, the field report and any potential NEA candidate are submitted by the reducer to the run coordinator who double-checks NEA candidates before submitting them to MPC. Later, the centralized field reports are submitted in batches to MPC. Upon source validation by the reducers, in the near future we plan to integrate in the NEARBY platform this entire chain to check and submit the MPC reports automatically. 

\subsection{NEARBY NEA Discoveries}
Our Nov 2018 run was affected about half time by bad seeing (above 2") and winds. Nevertheless during good seeing time (about 20 hours), this run actually produced the first two NEA discoveries credited to NEARBY, which we remind bellow and include in Figure~\ref{fig:discov}. 

\texttt{2018 VQ1 = E223100}

This object was discovered by the observer and reducer (MPEC 2018-X85) who spotted it as a quite faint (R=21.2) and relatively fast ($\mu$=4.1"/min) small trail which escaped the automate detection of NEARBY. This discovery happened in the E223 field observed in average seeing 1.8" during 1/2 Nov 2018 night (second of our run), during careful blinking in Astrometrica of the NEARBY reduced images. Thanks to the near real time reduction and visual quick-scan of the reducer, we submitted the MPC report in less than two hours after observing the field. Moreover, the fast reaction allowed us to recover the object two hours later, which was essential for second night recovery in very bad seeing conditions. 

\texttt{2018 VN3 = E522022}

This object was discovered by the NEARBY detection platform and was validated by the reducer (MPEC 2018-Y90) as a relatively bright (R=19.7) and slow speed ($\mu$=2.1"/min) star-like apparition. It happened in the E522 field observed in good seeing 1.5" during 5/6 Nov 2018 (the last night of our run), and it was reported only one and half hour after discovery, allowing us immediate recovery which enlarged the arc to two hours. 

\texttt{2018 VE4 = E165100}

Another NEA first imaged by the INT was this very fast object ($\mu$=11.7"/min) relatively faint (V$\sim$20.4) in the E165 field taken at 1 Nov 2018 (our first night) in bad seeing (2.5") around 2 UT. The reducer spotted it as a faint trail next morning, and we submitted the report 7 hours after the observation. The same field was imaged 5 hours after us by another survey (around 7 UT), being reported by their pipeline earlier than us, thus unfortunately we lost the discovery credit (MPEC 2018-Y44). 

Besides these 3 NEAs, other 18 faster objects were recovered and followed-up during the run, resulting in a few Mars-crosser asteroids discovered by NEARBY.

\begin{figure}
\centering
\includegraphics[width=0.95\linewidth]{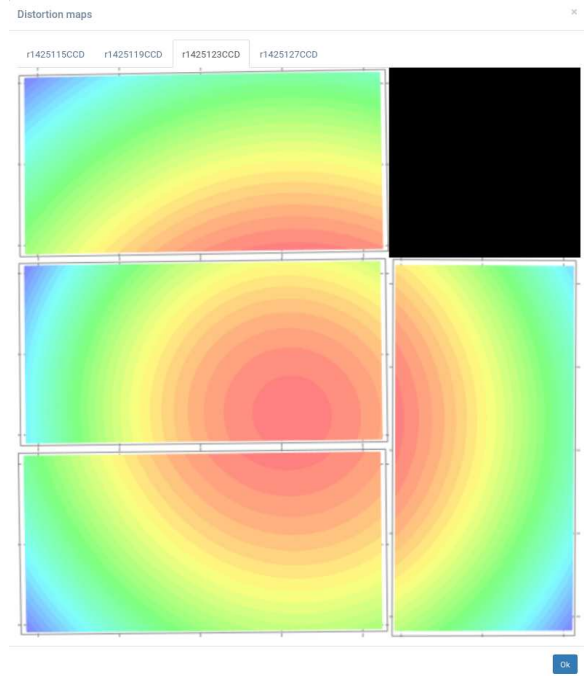}
\caption{Distortion map for the third image (r1425123.fit) of the field E557 observed at 5/6 Nov 2018 with the INT/WFC resolved with Scamp independently for each CCD and mosaicked by the NEARBY interface for checking by the assistant reducer.}
\label{fig:distortion}
\end{figure}

\section{Data Analysis}
We assess next the astrometry accuracy and the photometry limits of the NEARBY asteroid detections based on our INT-WFC mini-survey work.

\subsection{Field Correction}
Most larger field prime focus cameras (including WFC) suffer of image distortion effects which need to be corrected in order to achieve accurate astrometry across the whole observed field. In most cases Astrometrica could not accommodate the WFC distortion and in some cases the entire field recognition fails, especially in CCD3 which is more distant to the optical centre and affected by most bad columns. Therefore in our previous EURONEAR work, before Astrometrica we had to use THELI for image reduction and field correction. Upon automate image treatment in IRAF (bias, flat field and bad pixels), the NEARBY pipeline calls SExtractor, Scamp, and Swarp, which correct the field distortion using any reference catalog. Upon some testing using different catalogs, for our INT runs we decided to use 2MASS \citep{skrutskie06} which is homogenous across all sky and deep enough to provide a few dozens or hundred reference stars across each WFC CCD. For each field, the field recognition and image correction could be checked by the assistant reducer who can easily access the distortion field maps from the field interface (Figure~\ref{fig:distortion}).

\begin{figure}
\centering
\includegraphics[width=0.9\linewidth]{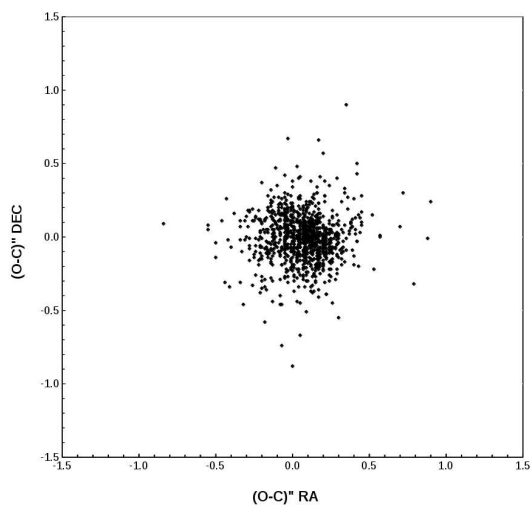}
\caption{(O-C) residuals for about 300 known numbered asteroids observed during 5/6 Nov 2018. 
The O-Cs were calculated using the FITSBLINK server \citep{skvarc18}.}
\label{fig:ominc}
\end{figure}

\subsection{NEARBY Astrometric Accuracy}
We could assess the NEARBY astrometric accuracy by plotting the observed minus calculated (O-C) positions of the known numbered asteroids, which have very good accuracy. In Figure~\ref{fig:ominc} we present an example using data for 60 WFC fields observed in 5/6 Nov night in average seeing 1.5-2.5". More than 1200 measurements for about 300 numbered known asteroids as faint as V$\sim$22 are plotted, showing average total residuals O-C=0.21" (0.06" in $\alpha$ and 0.01" in $\delta$).

\begin{figure}
\centering
\includegraphics[width=1.0\linewidth]{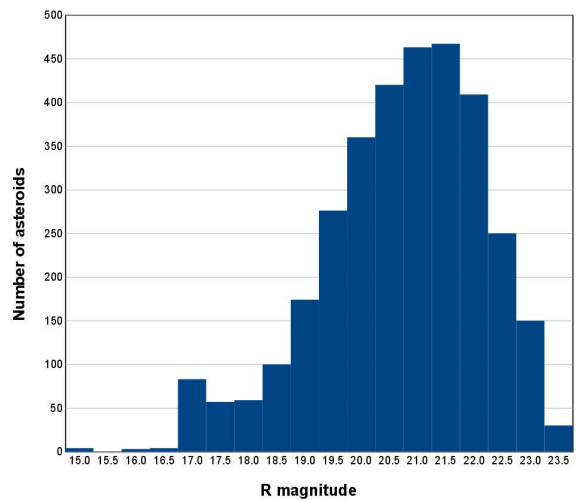}
\caption{Sample of histogram plotting the INT photometric survey limits 
for valid asteroid NEARBY detections (70 WFC fields observed during 2/3 Nov 2018 night).}
\label{fig:histo}
\end{figure}

\subsection{NEARBY INT/WFC Photometry Limits}
Figure~\ref{fig:histo} plots the histogram of valid moving objects function of their apparent magnitude, based on all our INT 2/3 Nov 2018 data representing more than 3300 measurements for 800 asteroids observed in 70 fields with 30s exposure time in average seeing conditions (1.5"-2.5"). An apparent maximum around R=21.5 is visible, with few hundred asteroids detected to R=23 and few dozen up to R=23.5 limiting magnitude.

\section{Conclusions and Future Work}
Thanks to the grant from the Romanian Space Agency (ROSA by ESA-SSA segment), in 2018 the NEARBY platform aiming for automatic detection of asteroids and near Earth asteroids (NEAs) has been written and deployed at Technical University of Cluj-Napoca and tested at the University of Craiova, Romania, based on know-how and data made available by the ING and IAC.

The NEARBY pipeline builds on the previous EURONEAR community experience and network access to medium-size telescopes in Europe and Chile. In this sense, in 2017-2018 we used the Isaac Newton Telescope (INT) and the Wide Field Camera (WFC) for archival and real time testing during a few pilot mini-surveys (including the largest one of 5 nights in Nov 2018). 

NEARBY seems better than other similar asteroid detection software such as Astrometrica (by about 10\%) but its detection rate remains bellow human detection (by about 10\%). Compared with our past work, NEARBY decreases data reduction time from three hours to half hour for each observed INT/WFC field.

Few functions are planned to be implemented soon in the NEARBY platform with the aim to ease the survey work (of the observers and run coordinator) using the INT and other larger telescopes and cameras feasible for larger surveys. The most important tasks under work are the followings: 

\begin{itemize}
\item Implement the image bulk upload, more fields or survey area, and image sorting on the server, instead of the actual upload field by field and sorting by the observer. 
\item Implement custom user-friendly camera definition using instrument configuration file, which defines the geometry camera by a few parameters (e.g. pixel scale, binning, essential header keywords, CCDs, sizes and positions relative to the camera center). In this sense, we will test first data acquired in March 2018 with the OGLE-IV camera (32 CCDs covering 1.3 sq. deg) mounted on the Warshaw 1.3m telescope in Chile. 
\item Implement automate NEARBY image binning mode in case of bad seeing or use of small pixel size cameras, in order to increase the signal to noise of the faint objects and minimize the trail loss effect of faster NEAs. 
\item Unify all detection parameters based on the actual seeing, measured by NEARBY in the current image, and observing sequence (e.g. exposure time, cadence, sequence and typical MBA proper motion in the observed field). 
\item Integrate on the NEARBY platform automatic object identification, quality control tests, NEA candidate alerts and immediate MPC report submission once a field has been validated by the experienced assistant reducers.
\end{itemize}

\section*{Acknowledgments}

This research has been supported by ROSA (Romanian Space Agency) by the Contract CDI-STAR 192/2017, NEARBY - Visual Analysis of Multidimensional Astrophysics Data for Moving Objects Detection. 

The data used to test the NEARBY platform was available by observations made with the Isaac Newton Telescope (INT) operated on the island of La Palma by the Isaac Newton Group (ING) in the Spanish Observatorio del Roque de los Muchachos (ORM) of the Instituto de Astrofisica de Canarias (IAC). The authors acknowledge the Spanish and IAC time allocation committee which granted the INT/WFC observing time (programs C85/2018A, C9/2018B and C10/2018B, some service D-time and other past programs) which allowed us to test and assess NEARBY.

The following reducer assistants were involved in NEARBY and Astrometrica tests and field validation: Costin Boldea, Afrodita Boldea, Marian Predatu, Adrian Stanica, Viktoria Pinter and Alin Buhulea (University of Craiova), Elisabeta Petrescu, Daniel Bertesteanu, Andra Stoica, Andreea Timpea, Simon Anghel, Malin Stanescu (Bucharest Astroclub), Ruxandra Toma (Romanian Astronomical Institute) and Marcel Popescu (IAC and co-observer of the Nov 2018 run). The following amateurs and students reduced in Astrometrica some fields observed in 2014-2017 used for comparison with NEARBY: Lucian Hudin, Alex Tudorica, Adrian Sonka, Stefan Mihalea (Romania) and Farid Char (Chile). The following ING students were involved INT observing (2017-2018) and past data reduction related to this work: Thomas Davison, Thomas Wilson, Tarik Zegmott, Teo Mocnik and Vlad Tudor.

\end{document}